\documentclass[a4paper,11pt]{article} \usepackage[latin1]{inputenc}
\usepackage[T1]{fontenc}
\usepackage{amsmath}
\usepackage{amsfonts}
\usepackage{amssymb}
\usepackage{color}
\usepackage{graphicx}

   \newcommand{\be}{\begin{equation}}
	 \newcommand{\ee}{\end{equation}}
	 \newcommand{\ba}{\begin{eqnarray}}
		 \newcommand{\ea}{\end{eqnarray}}
		 
		  \newcommand{\DD}{\!\!\not \!\!
		 D} \newcommand{\bea}{\begin{eqnarray}}
			 \newcommand{\eea}{\end{eqnarray}}

\begin{document} \title{Fermions in an AdS$_3$ Black Hole Background and
the Gauge-Gravity Duality}
\author{C.~D.~Fosco$^a$, E.~F.~Moreno$^b$  and
F.~A.~Schaposnik$^c$\thanks{Also at CICBA.} \\ \vspace{0.2 cm} \\
{\normalsize \it  $^{a}$Centro Atómico Bariloche and Instituto Balseiro}\\
{\normalsize \it  Comisión Nacional de Energía Atómica}\\ {\normalsize \it R8402AGP Bariloche, Argentina}\\
{\normalsize \it $^b$Department of Physics, Northeastern University}\\ {\normalsize \it Boston, MA 02115,
USA.} \\
{\normalsize \it $^c$\it Departamento de F\'\i sica, Universidad Nacional de La Plata}\\ {\normalsize \it Instituto de F\'\i sica La Plata}\\ {\normalsize\it C.C. 67, 1900 La Plata, Argentina}}

\date{\today}

\maketitle
\begin{abstract}
We study a model whose dynamics is determined by a Maxwell Lagrangian coupled
to a complex scalar and a Dirac fermion field, in an $AdS_3$ black hole
background.  Our study is performed within	the context of the Euclidean
formalism, in terms of an effective action $S^{eff}$ that results from
integrating out the fermion	field.  In particular, $S^{eff}$ includes an
induced parity breaking part which reduces, in the weak coupling limit, to
Chern-Simons terms for both the gauge and spin connections, with temperature
dependent coefficients.  We find numerically the effective action
minimum and, applying the AdS/CFT correspondence, we discuss the
properties of the dual quantum field theory defined on the boundary.  We show
that, in contrast with what happens in the absence of fermions, the system does
not undergo a phase transition at any finite temperature.
\end{abstract}
\section{Introduction}

A startling consequence of the parity anomaly~\cite{DJT} for three-dimensional
gauge theories coupled to fermion matter fields is that, because of the virtual
fluctuations the latter, a Chern-Simons (CS) action $S_{CS}[A]$ for the  gauge
field connection $A$~\cite{Red} may be induced.  This result has been obtained
by following different approaches; among them functional methods~\cite{RJH}, by
evaluating  the Atiyah-Patodi-Singer-$\eta$ invariant associated to the Dirac
operator~\cite{AG}, and by using a $\zeta$-function regularization for its
determinant~\cite{GRS}.  In an entirely similar fashion, a $2+1$ dimensional
Dirac field coupled to a gravitational field background, induces a
gravitational Chern-Simons term, an object which may be written in terms of the
spin connection $\omega$, $S_{CS}[\omega]$ or, alternatively, as
$S_{CS}[\Gamma]$ for the corresponding Christoffel connection
$\Gamma$~\cite{AG}, \cite{GS}-\cite{Kulikov}.

The finite temperature extension of these calculations is more involved and
leads to a nonlocal effective action~\cite{DGS}-\cite{FRS2}. Nevertheless, in
the weak coupling limit one still does get, for gauge theories, a CS term with
a temperature dependent coefficient.

The parity anomaly in odd-dimensional space has been exactly evaluated for
situations where both gauge and gravitational background fields are
present~\cite{AG}. However, to our knowledge, the form of the full (anomalous)
parity-odd contribution to  the fermion determinant  has not been calculated,
even at zero temperature. As we shall show, this term becomes just the sum of
two CS terms, one for the gauge field and the other for the spin connection:
\begin{equation} \left.\log\det (\; \DD[A,\omega] + M) \right|_{odd} = i
	S_{CS}^{gauge}[A]\,+\, i S_{CS}^{grav}[\omega] \;.  \label{rhs0}
\end{equation}
This result   is relevant in connection with  applications of the AdS/CFT holographic correspondence to the study of condensed matter systems (for a complete list of references, see, for example, the reviews~\cite{cm1}-\cite{cm2}). In fact, much of the work on this subject has been restricted to  leading order in semiclassical  calculations in the bulk, namely  to evaluating the path integral that defines the partition function $Z^{bulk}$ on its Euclidean saddle point, so that $Z^{bulk} \approx \exp(-S_E)$ with $S_E$ the on-shell action connected to the free energy of the field theory through the usual relation $F = -T\log Z^{bulk}$.  When fermionic matter is present,  its contribution is usually considered within the Thomas-Fermi approximation~\cite{cm2}. More recently studies of the effects of fermion loops and in general quantum fermions have   been presented \cite{HH}-\cite{Schd}.

Using the results on the $d=3$ fermion determinant described above we shall
study, within the gauge/gravity framework, a gravity system consisting of an
AdS$_3$ black hole coupled to a $U(1)$ gauge field, a complex scalar field and
a massive Dirac fermion. After integrating out the fermions we obtain a (bulk)
Euclidean effective action $S_E^{eff}$, which incorporates the full parity-odd
fermionic contribution to the bulk partition function. Afterwards, using  an
appropriate ansatz, we shall numerically solve the equations of motion
associated to $S_E^{eff}$ and then use the AdS/CFT correspondence to study the
resulting dual theory defined on the boundary. This will leads to the main
result of our work, concerning the thermodynamical properties of the
corresponding dual 2-dimensional theory.

The paper is organized as follows. We introduce in section 2 the model in the
bulk with dynamics governed by a Maxwell Lagrangian coupled to a charged scalar
and a Dirac fermion in an AdS$_3$ black hole background. Then we show, in
Sec.ion 3, that when both gauge and spin connections are taken into account
there are no mixing terms and the complete result is just the sum of the two
respective CS terms, as written in eq.(\ref{rhs0}). Having obtained the
effective action resulting from integration of fermions, we analyze in Sec.ion
4 its classical equations of motion and determine the appropriate boundary
conditions to apply the gauge/gravity correspondence. We then proceed  to solve
(numerically) the coupled system of equations and, from the asymptotic behavior
of the solutions, we analyze the critical behavior of the quantum field theory
defined in the boundary. Our results are discussed in Sec. 5.

section{The model} \subsection{Notation} The most relevant degrees of
freedom we shall consider correspond to a $U(1)$ gauge field connection
\be A \equiv A_\mu dx^\mu \ee
defined on a $d= 3$ manifold ${\cal M}$  (the bulk) with  $\mu = 1,2,3$.
The  2-form field strength $F$ is then given by
\be F = dA  = \frac12\left( \partial_\mu A_\nu  - \partial_\nu A_\mu
\right) dx^\mu \wedge dx^\nu \ee

\noindent The dreibein fields $\{e^a\}$ $(a=1,2,3)$ are an orthonormal
basis of one forms on ${\cal M}$, locally given by \be e^a = e^a_\mu dx^\mu
\;.  \ee

\noindent The spin connection, $\omega_{ab}$, is an $SO(3)$-valued 1-form on
${\cal M}$ satisfying
\be de_a + \omega_{ab}e_b = 0 \; , \;\;\;\; \omega_{ab}=-\omega_{ba} \; .
\ee
In local coordinates
\be \omega_{ab} = \omega_{ab}^\mu dx_\mu = \omega_{abc} e_c \ee
where
\be \omega_{ab\mu} = e_{a\nu} e^\nu_{\;\,b;\mu} \; , \;\;\;\; e_b^\nu =
g^{\nu\rho} e_{b\rho}\;.  \ee
The semicolon refers to differentiation using the Christoffel symbol $
\Gamma_{\nu\alpha}^\mu $,
\be v^\mu_{\;\, ;\alpha} = \partial_\alpha v^\mu + \Gamma^\mu _{\nu\alpha}
v^\nu \;, \ee
and, finally, the curvature 2-form $R$ associated to $\omega$ is given by
\be R_{ab} = d \omega_{ab} + \omega_{ac} \omega_{cb} \;.  \ee

\noindent Regarding Dirac field related conventions, we use an irreducible
representation of Dirac algebra, based on $2\times2$ $\gamma$-matrices
which satisfy
\[ \gamma^\mu(x) = e^\mu_a(x) \gamma_a \;,\;\; \gamma_a^\dagger =
-\gamma_a\;,\;\; \{\gamma_a,\gamma_b\} = 2 \delta_{ab} \;,\;\; \sigma^{ab}
= \frac{1}{4i} [\gamma^a,\gamma^b]\; \]
while for the spin connection we have
\be \omega_\mu(x)= \frac{1}{2}\omega_{\mu}^{ab}\sigma_{ab} \;.  \ee

\subsection{The black hole  background} We consider an anti-de Sitter
($2+1$)-dimensional Ba\~nados-Teitelboim-Zanelli  (BTZ) black hole \cite{BTZ}-\cite{BTZ2} in Poincaré
coordinates with Lorentzian signature $(-1,1,1,1)$, such that
\be ds^2 = \frac{L^2}{z^2} \left(-f(z) dt^2 + f^{-1}(z) dz^2 + dx^2 \right)
\;, \label{unoa} \ee
\be f(z) = 1 - \frac{z^2}{z_h^2} \;.  \label{dosa} \ee
Here, $L$ denotes the AdS radius, and $L^2 = -(1/\Lambda)$, with $\Lambda$
the negative cosmological constant. The boundary is at $z=0$ and the
horizon at $z=z_h$. The Hawking temperature becomes \be T =
\frac{|f'(z_H)|}{4\pi}  = \frac1{2\pi z_h} \;.  \ee

\subsection{The bulk action} We start from a  $U(1)$ gauge field with
Maxwell Lagrangian coupled to a charged scalar and a Dirac fermion in the
background of the black hole metric (\ref{unoa})-(\ref{dosa}). The action
takes the form
\be S = \int d^3x \sqrt{|g|} \left( -\frac1{4} F^{\mu\nu}F_{\mu\nu}
-\frac12 D^\mu \Phi^* D_\mu \Phi + \bar \psi \DD \psi + M \bar\psi \psi-
V(|\Phi|) \right) \label{action} \ee
Here, $e$ denotes the gauge coupling,  $\Phi$ is a complex scalar, and
$\psi$ a two-component Dirac fermion. The derivatives are defined,
respectively, as
\begin{align} &D_\mu \Phi = (\partial_\mu - ie A_\mu)\Phi \;, \nonumber\\
&\DD \psi=  \gamma^\mu\, \left(\partial_\mu \,+\,eA_\mu(x) \,+\,
\omega_\mu(x) \right) \psi \;.  \end{align}
Since in ($2+1$)-dimensional space $e^2$ has the dimensions of a mass, in order to work with dimensionless quantities, we redefine $e^2 \to e^2 L$ (a similar rescaling is done with all dimensionful quantities).  The form of $V(|\Phi|)$ will be made explicit below.

The partition function for the theory with dynamics governed by action
(\ref{action}), defined  in the black hole background
(\ref{unoa})-(\ref{dosa}), is then written as a functional integral
\be Z^{bulk} = \int DA_\mu D\Phi D\bar \psi D\psi\, e^{iS} \;.
\label{part} \ee
This partition function  plays a central role in the gauge-gravity duality:
its weak-coupling limit will give information on the strong coupling
behavior of the dual quantum field theory defined on the boundary.

Note that, within the gauge/gravity framework, one usually works  in the
classical approximation, writing $Z \sim \exp(i S_{cl})$ where ``$ cl\,$''
indicates that the bulk classical action is taken on-shell, with the
appropriate boundary conditions for the classical solution. As discussed in
the introduction, we shall instead proceed to integrate over fermions thus
generating an effective action that already includes fermionic quantum
effects in the bulk.  That is, we path-integrate over the fermions,
obtaining
\be Z = \int DA_\mu D\Phi  \; e^{iS^{eff}} \;, \label{part2} \ee
where
\be S^{eff} = \int d^3x \sqrt{|g|} \left(-\frac1{4} F^{\mu\nu}F_{\mu\nu}
-\frac12 D_\mu \Phi^* D_\mu \Phi - V(|\Phi|) \right)  -i \log\det (\;\DD +
M) \;.  \label{ef} \ee
\section{The fermion determinant} \subsection{The fermion determinant at
zero temperature}

To our knowledge, explicit results for the fermion determinant in
(\ref{ef}) have   been reported  when just one of the two connections is
present. We shall discuss here the case in which  fermions are coupled to
both $A$ and $\omega$ (or its associated $\Gamma$) showing that the result
becomes just the sum of the two separate CS terms.

We are interested in the induced parity-violating term, understanding by
such the parity-violating $3$-form in the effective action due to the
fermion loop contribution.

Although we are interested in the $U(1)$ gauge theory,  we shall also
consider in this Sec.ion the case of an $SU(N)$ gauge group since the
demonstration does not turn to be more complicated than in the Abelian
case. Concerning the gravitational CS term, as already mentioned, it can be
expressed either in terms of the spin or the associated Christoffel
connection depending on the anomaly calculation one chooses (no coordinate
anomaly or no frame anomaly respectively).  For definiteness we shall
choose the Christoffel connection $\Gamma$ to write the induced parity-odd
term.

We start by separating out the parity odd and even contributions to the
fermion determinant,
\be \log\det (\; \DD + M) =\left.\log\det (\; \DD + M) \right|_{even} +
\left.\log\det (\; \DD + M) \right|_{odd}.
 \ee
For the case of purely $U(1)$ gauge theory (no spin
connection), and keeping the leading terms in the IR limit, the even part
takes the form~\cite{DJT}-\cite{GS}
\begin{align}
\left.\log\det (\; \DD + M) \right|_{even}[A] =& \frac{i e^2}{48\pi |M|} \int\sqrt
{|g|}d^3x \left(F_{\mu\nu}^2 \vphantom{\frac{1}{60}} \right.\nonumber\\
& +\left. \frac{i}{60 M^2}\left(
18F^{\nu\lambda}F_{\nu\lambda,\mu}^{\;\;\;\;\;\; \mu} + 7F^{\nu\lambda,\mu}F_{\nu\lambda,\mu} - 2F^{\nu\lambda,}_\lambda F_{\nu\mu,}^{\;\;\;\; \mu}
\right)\right)
\end{align}
We shall however discard this even contribution, since we will consider from now
on  the large $M$ limit; the same holds true for the gravitational contribution.

~%

By a perturbative expansion argument, it is evident that $S_{odd}$ will
contain -at least- the purely gauge and the purely gravitational CS terms
found in \cite{Red},\cite{AG},\cite{Ojima}-\cite{Kulikov}
\begin{equation}
\left.\log\det (\; \DD + M) \right|_{odd} = i S_{CS}^{gauge} \,+\,
i S_{CS}^{grav}\,+\,iS^{mixed}\;
\label{rhs}
\end{equation}
where
\begin{equation}
S_{CS}^{gauge}\,=\, \frac{\alpha e^2}{16 \pi} \frac{M}{|M|}\, \int d^3x \, {\rm tr} \big(A \wedge
dA + \frac{2e}{3} A \wedge A \wedge A\big) \;,
\label{20}
\end{equation}
with $\alpha= 2, 1$ for $U(1),SU(N)$ respectively and
\begin{equation}
S_{CS}^{grav}\,=\, \frac{1}{32\pi} \frac{M}{|M|}\, \int d^3x \, {\rm tr} \big(
\Gamma \wedge d \Gamma + \frac{2}{3} \Gamma
\wedge \Gamma \wedge \Gamma\big) \;.
\label{21}
\end{equation}
The term $S^{mixed}$ represent possible  contributions mixing $A$ and
$\Gamma$. We will show that it vanishes.
We use the standard $1$-form matrix notation for both terms; for example: $\Gamma = \Gamma_\mu dx^\mu$, where $\Gamma_\mu$ is, for each $\mu$, a $3\times 3$ matrix: $(\Gamma_\mu)^\alpha_\beta =
\Gamma_{\mu\beta}^\alpha$.
For the gauge field, we have, in the case of a  gauge group which can be $SU(N)$ or $U(1)$  (with generators $t^A$),
$A = A_\mu dx^\mu$, where $A_\mu=A_\mu^B t^B$.
The $t_r$ traces refer, in each cases, to the  corresponding product of matrices.

Using again perturbative arguments, let us note that
\begin{align}
S_{odd} &=  i\lim_{M\to\infty} {\rm Im}
\big[\log \det(\not\!\!D +M) \big]
\nonumber\\
&= i\lim_{M\to\infty} {\rm Im} \Big\{
-\frac{1}{2}
{\rm Tr} \big[(\not\!\partial + M)^{-1}(\not\!\!A +\not\!\!\omega)\big]^2
+\frac{1}{3}
{\rm Tr} \big[(\not\!\partial + M)^{-1}(\not\!\!A +\not\!\!\omega)\big]^3
\Big\}
\end{align}
where the $T_r$  is a functional and internal space trace and we have
discarded terms that cannot produce $3$-forms.

Because of the one to one relation between $\omega$ and $\Gamma$, the
expression above shows explicitly that { $S_{mixed}$ can only be
constructed in terms of the $1$-forms $A$, $\Gamma$, and the exterior
derivative $d$}. Since we are taking the large $M$ limit, the leading order
terms will have only three of them.
There are two possible cases:
\begin{enumerate}
\item Terms involving one exterior derivative, $d$.
Any $3$-form must, in this case, contain one $\Gamma$ and one $A$
(otherwise it would not be part of the mixed term). Thus, it should
be the integral of a term of the form ${\rm tr} (\Gamma \wedge  d A)$ or
${\rm tr} (A \wedge d \Gamma)$. We should distinguish between the non Abelian and Abelian
cases:
\begin{enumerate}
\item Non Abelian gauge field case: it is not possible to build such a
term, since the trace of $A$ is zero.
\item Abelian gauge field case: there is no trace over the Lie algebra, and
a careful examination shows that, in a torsion-free spacetime, the only
possible non vanishing term is of the form:
$$\Gamma^\alpha_{\mu\alpha} \epsilon_{\mu\nu\rho} \partial_\nu
A_\rho$$
which is a total derivative, since:
\begin{equation}
\Gamma^\alpha_{\mu\alpha} =
\frac{\partial\log(\sqrt{g})}{\partial x_\mu} \;,
\end{equation}
and we assume $\epsilon_{\mu\nu\rho} \partial_\mu\partial_\nu A_\rho = 0$
\end{enumerate}
\item Terms without any exterior derivative.
In this case $3$-forms contributing to the mixed term have to be of the form
$\Gamma \wedge A \wedge A$ or $A \wedge \Gamma \wedge \Gamma$
(or permutations of them). Now
\begin{enumerate}
\item The first kind of term (two $A$'s) vanishes, because after contracting the ${\cal G}$-algebra indices, we end up with the three objects:
$A_\mu^B A_\nu^B$, $\epsilon^{\alpha\beta\gamma}$ and
$\Gamma_{\rho\nu}^\lambda$, which should be contracted and balanced.
The two $A$'s form a symmetric tensor in $\mu,\nu$. Thus they cannot be
contracted both with $\epsilon$. One of them has to be contracted with
$\epsilon$, though (otherwise one should contract two indices in
$\epsilon$).

Then we are left with two indices in
$\epsilon$ to be contracted with $\Gamma_{\rho\nu}^\lambda$ . This yields
zero, in a torsion-free spacetime.

The argument above, with a minimal modification, holds true in the Abelian
case.
\item The second kind of term (two $\Gamma$'s) vanishes since the trace
over the Lie algebra is zero (there is only one $A$).
In the Abelian case, it is possible to balance the indices; however, such
an object would break Abelian gauge invariance, since it would be necessary  of the
form:
$\epsilon^{\mu\nu\lambda} \Gamma^\beta_{\mu\alpha}
\Gamma^\alpha_{\nu\beta} A_\lambda$.
\end{enumerate}
\end{enumerate}
Having enumerated all the possible nontrivial options, we conclude that
\be
S_{mixed}=0
\label{corta}
\ee.

\noindent Remark: strictly speaking, it is sufficient to prove just 1 (absence of
terms with a derivative), since it is not possible to construct a general
covariant scalar just involving connections.

In summary, we see from  eq.(\ref{rhs}) that the fermion determinant is just the product  of the gauge and gravitational determinants
\begin{equation}
\left. \log \det (\; \DD + M) \right|_{odd} =  i \left( S_{CS}^{gauge} \,+\,
 S_{CS}^{grav}\right)\
\label{rhsfinal}
\end{equation}

\subsection{The fermion determinant at finite temperature}

The parity-violating part of the finite temperature fermion determinant for the case in which
there is only a  $U(1)$ gauge connection has been computed in Refs. \cite{DGS}-\cite{FRS2}.
The answer may be put in the form
\be
S_{odd}[A] = i\frac{e}{2\pi}\int d^2x \varepsilon_{jk}\partial_j A_k
\arctan\left( \tanh\left(\frac{\beta M}{2}\right) \tan
\left(\frac{e}{2} \int_0^\beta d\tau A_3
\right)
\right) \; , \;\;\;\; j,k=1,2 \;.
\label{eqjulio}
\ee
Here, the odd-parity action $S_{odd}$  is defined in Euclidean time $\tau
\in (0,\beta  = 1/T)$.

An expansion of (\ref{eqjulio}) in powers of $e$ yields the perturbative result
\begin{align}
S_{odd}[A]
 &= \frac{i}2 \tanh\left(\frac{\beta M}2\right)
 \frac{e^2}{4\pi}\int d^3x \varepsilon_{\mu\nu\alpha}A_\mu\partial_\nu A_\alpha
 + O(e^4)
 \end{align}
Following the approach in \cite{FRS1}-\cite{FRS2} one has an
analogous result for the case of the spin conection. The
arguments leading to zero mixed term still hold so that the
result for the complete odd-parity contribution to the finite
temperature fermion determinant is
\be S_{odd}[A,\Gamma;T] = {i} \tanh\left(\frac{\beta
M}2\right)\left( S_{CS}^{gauge}[A] +  S_{CS}^{grav}[\omega]
+O(e^4) \right) \label{SSC} \ee
where the Chern-Simons actions adopts the form
\begin{align}
 S_{CS}^{gauge}[A] &=  \frac{e^2}{8\pi}\int d^3x
\varepsilon_{\mu\nu\alpha}A_\mu\partial_\nu A_\alpha
\nonumber\\
 S_{CS}^{grav}[\omega] &= \frac{1}{32\pi}\int d^3x \epsilon_{\mu\nu\alpha}\left(
(\partial_\mu \omega_{a \nu}^{\;b}) \omega_{b \alpha}^{\; a} +
\frac23 \omega_{a \nu}^{\,b} \omega_{b \mu}^{\,c} \omega_{a
\alpha}^{\,a} \right) \;.
\label{SSCC}
\end{align}
\section{The effective action and the AdS$_3$/CFT$_2$
correspondence}
We shall follow the AdS/CFT approach to the  study of a
strongly coupled system   defined on the boundary of an
asymptotically AdS space (the bulk).  If the bulk metric
corresponds to a black hole with Hawking temperature $T$
then, according to the AdS/CFT correspondence,  the dual theory in
the boundary is a quantum field theory at finite temperature
$T$ and its properties should be studied by performing a
semiclassical approximation of the (Euclidean) partition
function in the bulk, $Z^{bulk}$.

The general form of the AdS/CFT conjecture reads
\be Z^{bulk}[\Sigma \to \Sigma_0] = \left\langle
\exp\left(-\int_{boundary} \hspace{-1 cm} d^2x \, \Sigma_0
\,{\cal O}~~\right)\right\rangle \label{Z} \ee
where
\be Z^{bulk}[\Sigma \to \Sigma_0] \equiv  \int_{\Sigma =
\Sigma_0} D\Sigma \exp\left( -S[\Phi] \right) \;.
\label{adscft}
\ee
Here, $Z_{bulk}$ represent the generating functional for the theory in the
bulk, which in the present case corresponds to asymptotically $AdS_3$
spacetime. We denote with $\Sigma$ all the fields in the bulk which tend on
the boundary  to the value $\Sigma_0$  up to an overall power of $z$.
Concerning  ${\cal O}$, it  denotes the operators for the quantum field theory in the
boundary associated to each field in $\Sigma$ .

Usually, one takes the classical limit in the r.h.s. of
(\ref{adscft}) in which case the path integral  defining $Z^{bulk}$
is approximated by $\exp(-S_\text{on shell}$). In the present
case we have gone one step further and integrated out fermions,
what lead us to the effective action (\ref{efectiva}).
This yields an effective action with a parity violation
term which, in the weak coupling regime corresponds to a Chern-Simons term.
Now, this term arises as a phase
both with Euclidean and Lorentzian metric signature (a well-known fact resulting from the topological character of CS
terms).  As a result, the Euclidean effective action has complex saddle points which,
nevertheless, have proven to be of relevance  in problems like, for
example,  that of  confinement in compact $QED_3$ \cite{FS}-\cite{Hos}.
Here we shall look for complex saddle points of $S^{eff}_E$   leading to a
real on-shell Euclidean action and, consequently, to a sensible free energy
$F = - T S^{eff}_E$.

We shall then start, for the study of the correspondence, from
\be
Z^{bulk} = \int DA_\mu D\Phi \exp\left(-S_E^{eff}\right)
\label{vos}
\ee
where the Euclidean version of the effective action $S_E^{eff}$  is
obtained inserting (\ref{SSC})-(\ref{SSCC}) in the
 effective action $S^{eff}$  (\ref{ef}) and then  Wick rotating,
\begin{align}
S_E^{eff} =& \int d^3x \sqrt{|g|} \left( -\frac1{4}
F^{\mu\nu}F_{\mu\nu} -\frac12 D_\mu \Phi^* D^\mu \Phi -
V(|\Phi|)
\right) \nonumber\\
& -\frac{i}2 \tanh\left(\frac{\beta M}2\right)\left(
\frac{e^2}{4\pi}\int d^3x \varepsilon_{\mu\nu\alpha}
A_\mu\partial_\nu A_\alpha + \frac{1}{32\pi}\int d^3x
\epsilon_{\mu\nu\alpha} (\partial_\mu \omega_{a \nu}^{\;b})
\omega_{b \alpha}^{\; a} + \frac23 \omega_{a \nu}^{\,b}
\omega_{b \mu}^{\,c} \omega_{a \alpha}^{\,a}
\right)\;.
\label{efectiva}
\end{align}
%
One then has
\begin{eqnarray}
Z^{bulk} &=&
\exp\left(-\int d^3x \sqrt{|g|}
\left(-\frac1{4} F^{\mu\nu}F_{\mu\nu}
-\frac12 D_\mu \Phi^* D^\mu \Phi
- V(|\Phi|) \right)\right.
\nonumber\\
&& \times \exp\left( \left.-\frac{i}2 \tanh\left(\frac{\beta M}2\right)
\left( \frac{e^2}{4\pi}\int d^3x \varepsilon_{\mu\nu\alpha}
A_\mu\partial_\nu A_\alpha
\right)
\right)
\right) \;.
\end{eqnarray}
We are not including the contribution of the spin connection
Chern-Simons action since we are taking the AdS black hole as a
background metric so that its contribution is just an
irrelevant factor both for computing vacuum expectation values of operators for
the theory on the boundary and also for differences of free energies
(see below).

Taking an AdS$_3$ black hole as background metric and making for the gauge
and scalar field the ansatz:  $A_0=\phi(z)$, $A_x=i\, A(z)$ and $\Phi =
\psi(z)$, the equations of motion associated to the effective action
(\ref{efectiva}) become
\begin{align}
&\psi'' - \frac{(1+z^2)}{(1-z^2)}\frac{1}{z}\, \psi' +
\left(-\frac{z_h^2 \,\phi^2}{(1-z^2)} - \frac{m^2}{z^2} +
z_h^2\,A^2\right) \frac{\psi}{1-z^2}=0 \nonumber\\
&\phi'' + \frac{\phi'}{z} - 2\frac{\psi^2}{z^2(1-z^2)}\phi
-\frac{2 \kappa}{z} A'=0 \nonumber\\
&
A''+\frac{1-3z^2}{1-z^2}\, \frac{1}{z} A' -
2\frac{\psi^2}{z^2(1-z^2)} A -\frac{2\kappa}{z(1-z^2)}\phi'=0
\label{sistema}
\end{align}
with
\[\kappa=\frac{e^2}{8\pi} \tanh(\frac{M}{2 T})\]
where we have rescaled $z\to z\, z_h$ so the horizon is located at $z=1$ and we have taken $V(|\Phi|) = m^2\, |\Phi|^2 $.

Concerning the plain symmetric ansatz chosen for the scalar and gauge fields, it is the simplest one leading to non-trivial results in models like the one we are discussing. Indeed, just by considering that the field dependence on coordinates reduces to the $z$ variable which goes from the horizon to the boundary allows one to establish the holographic correspondence without the complication introduced by considering more involved ansatz (see \cite{cm1}-\cite{cm2} for a more extended discussion on this
issue).

Concerning the scalar mass $m^2$, stability studies of Maxwell-scalar field models in an AdS black hole background have shown that, for all masses satisfying the Breitenlohner-Friedman (BF) bound $m \geq m_{BF}$,  there is no drastic change in the behavior of the system \cite{HR}. The main feature concerning mass dependence is that as  the mass grows, it is harder for the scalar hair to form, as it is to be expected. From now on we take the scalar mass $m$ in the Breitenlohner-Freedman bound, $m^2=-1$, a choice that will allow  comparison with the results in \cite{Ren:2010ha} obtained for the case in which fermions -and hence the Chern-Simons term- are absent.

Concerning boundary conditions, for $z \sim 0$ we have
\begin{align}
&\psi=D_1\, z\, \log z + D_2\, z\label{Des}\\
&\phi=B_1 + B_2 z^{2\kappa}  + B_3 z^{-2\kappa}   \\
&A=C_1 + B_2 z^{2\kappa}  - B_3 z^{-2\kappa} \label{Dos}
\end{align}
with $D_i, B_i, C_1$ real constants. As for the solution at the horizon $z=1$
 {\begin{align}
&\phi(1)=0\label{bc11}\\
&2\psi'(1) - \left( {A(1)^2}\,{z_h^2} - m^2\right)\psi(1)=0\label{bc12}\\
&A'(1) + \psi(1)^2 A(1) + \kappa \phi'(1)=0 \;.
\label{bc13}
\end{align}}
We assume that  the solutions are analytic at the horizon so
that $\phi$ should vanish as   $1-z$.

Let us concentrate in the scalar field $\Phi= \psi(z)$ with
boundary conditions (\ref{Des}). Within the AdS/CFT approach
there are in principle  two possibilities to relate  the
associated scalar operator $O_i$ ($i=1,2$) for the quantum
field theory on the boundary with the asymptotic values of
$\psi$ :
\be \langle O_i\rangle = D_i \; ,  \hspace{2 cm}
\varepsilon^{ij}D_j = 0 \label{boundary} \ee
One has to decide which of the two possible boundaries should be
set to zero, and we shall choose  the most divergent one, which means that we set $D_1 = 0$ so that  $D_2$ will give the
expectation value. One can see that the alternative choice
(that is, taking the coefficient of the $z\log z$ term as the
expectation value) leads to a negative  Euclidean action so
that it should be discarded as can be seen by calculating the associated free energy (see the discussion in  the next Sec.ion).

A condensate phase would correspond to a     nontrivial
v.e.v.,  $ \langle O_2\rangle =  D_2   \ne 0$. This can be
tested by finding, numerically,  solutions to  the  boundary
value problem (\ref{boundary}), thus determining the
coefficients $D_2$ as a function of $T$ and the parameters of
the theory.
{ Concerning the boundary conditions for the gauge fields, the choices $B_2=0$ and $B_3=0$ give rise to virtually
the same Higgs configuration, so the condensate behavior is insensible to this election (see Fig. 1). For definiteness we choose the condition $B_2 = 0$ so that the
boundary conditions (\ref{Des})-(\ref{Dos}) reduce
to
\begin{align}
&\psi \sim  D_2\, z\label{DesR}\\
&\phi \sim B_1 +    B_3 z^{-2\kappa}\label{DisR}   \\
&A\sim C_1    - B_3 z^{-2\kappa} \;.
\label{DosR}
\end{align}
}
\newpage
\vspace{1 cm}
\begin{center}
\includegraphics[width=6 in]{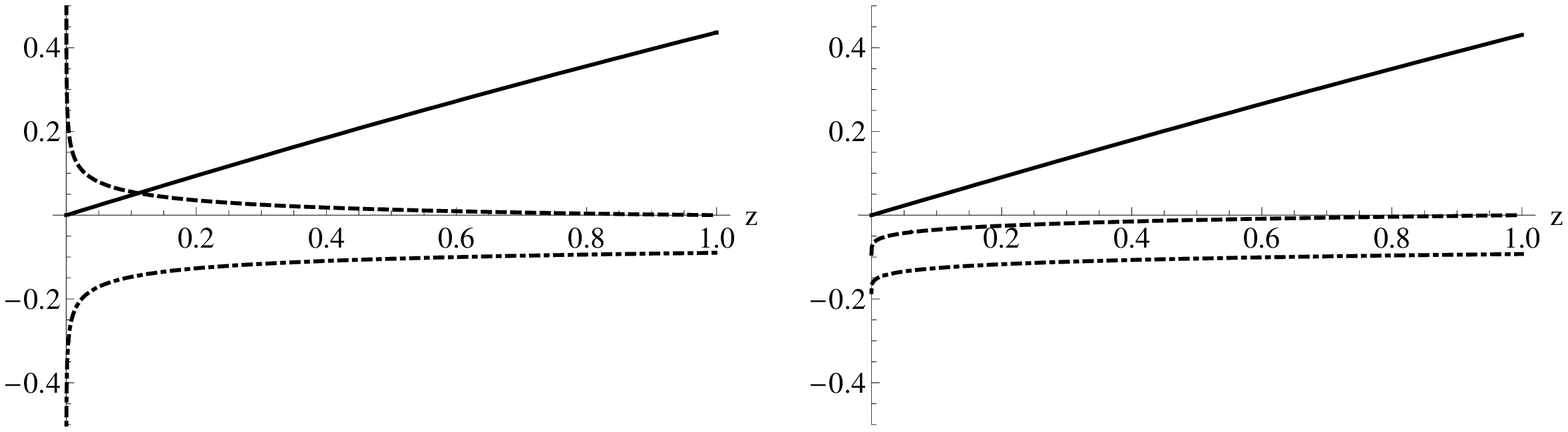}
\end{center}

\noindent Figure 1: Field profiles ($\psi$ - solid line, $\phi$ - dashed line, $A$ - dot-dashed line) for boundary conditions $B_2=0$ (left plot) and $B_3=0$ (right plot). Notice that, even though the gauge fields significantly differ in both cases, the Higgs profiles are essentially identical.

\vspace{1 cm}

Here $B_3$ should be identified with the chemical potential
and $B_1$ with the charge density.

Regarding the differential equations, we have employed a shooting method to
numerically solve them. The equations are integrated back from the horizon
$z=1$ with the initial conditions \eqref{bc11}-\eqref{bc13}. The solutions
depend on three parameters, that can be chosen to be $\psi(1)$, $\phi'(1)$,
and $A(1)$ respectively. These parameters are then adjusted (using, for
example, Newton-Fourier's method) so that the solutions satisfy the
conditions~\eqref{DesR}-\eqref{DosR} at the boundary $z=0$.

{  We numerically computed the  $D_2$ coefficient in the scalar field $\psi$
asymptotic expansion  as a function of the temperature $T$ for various values of  the CS coupling $\alpha$}. In contrast
with what happens when the CS term is absent (i.e., when fermions are not
present) the nontrivial solution exists at all temperatures.  In the
absence of fermions it has been shown in
\cite{Maity}-\cite{Lashkari:2010ak} that there exists a critical
temperature such that as $T \to T_c$ the condensed phase approaches the
normal (disordered phase) phase with $\psi = 0$. The presence of the
Chern-Simons term in the effective action (\ref{efectiva}) prevents this from
happening since, due to the Chern-Simons coupling between $A_0$ and $A_x$,
Eqs. (\ref{sistema}) had no nontrivial solution for a normal
(disordered) phase which would correspond to an  $A_0(z)= \phi(z) $ with
asymptotic behavior (\ref{DisR}) together with $\psi(z) = 0$ and $A_x(z)=
0$.

In connection with the fact that in the normal phase, the presence of the Chern-Simons term forces the gauge field to vanish, it is interesting to recall the results presented in \cite{Banados} concerning charged black holes in topologically massive electrodynamics. In particular, in this work it is shown that when a Chern-Simons term is present the theory cannot support a local electromagnetic field so that the solution of the Einstein-Maxwell-Chern-Simons equations corresponds to an uncharged (BTZ) black hole, thus forcing $A(z)$ and $\phi(z)$ to be zero. Interestingly enough this is true for any value of the Chern-Simons coefficient $\alpha=e^2/4\pi M \ne 0$ while for $\alpha = 0$ there is a nontrivial solution of the (2+1) Einstein-Maxwell equations corresponding to a charged black hole and a a charged gauge field, $\phi(z) \propto \log z$.

Remarkably, our numerical results go in the same direction as those in \cite{Banados} in the sense that they show  that the absence of a critical temperature seems to be present for any value of the CS parameter $\alpha$, so the transition to $\alpha \to 0$ (absence of CS term) is singular. We show in Fig. 2 the behavior of $D_2$ as a function of $T$ for three  values of  $\alpha \ne 0$ and we see that even for $\alpha$ as small as $ 0.001$ the behavior is radically different from what one find for $\alpha$ strictly zero.

\vspace{1 cm}
\begin{center}
\includegraphics[width=4 in]{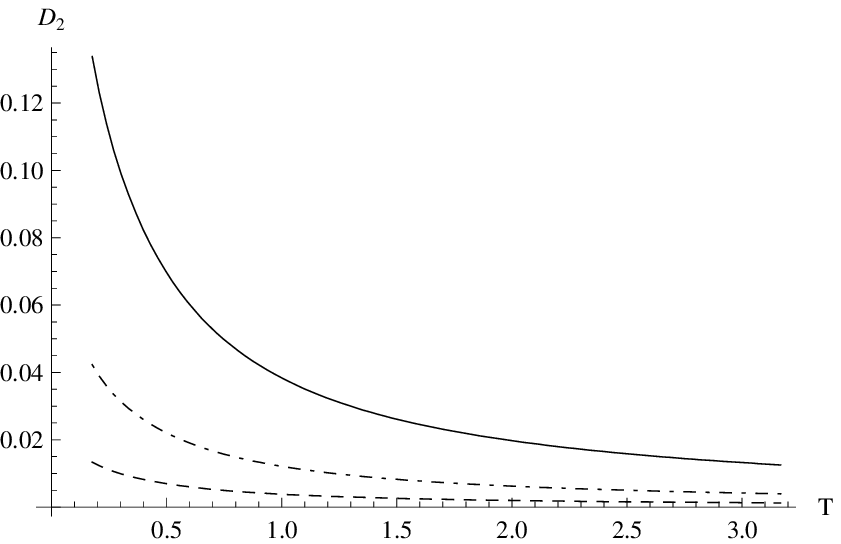}
\end{center}

\noindent Figure 2: The  $ D_2$ coefficient in the scalar field $\psi$ asymptotic expansion (with $D_1=0$) as a function of the temperature $T$ (measure in units of $M$) for $\alpha = (e^2/4\pi M) = 0.1$ (solid line), $\alpha=0.01$ (dot-dashed line), and $\alpha = 0.001$ (dashed line). In all cases $B_3 =0.1 $.

\vspace{1 cm}

As usual for this kind of system, the differential equations with  boundary
conditions \eqref{Des}-\eqref{bc13} support several solutions, with
an increasing number of nodes.
The free energy of each solution increases with
the number of nodes so it is  the solution with no nodes   the   thermodynamically
favored one. This behavior can be understood by noting that within the gauge/gravity
approach,the equation for the scalar field  in the bulk  can be view as a Schr\"odinger equation
in one spatial dimension where the well-known node theorem holds  \cite{Marc}.

\section{Free energy}
Within the gauge/gravity duality the free energy ${ F}$ of the theory
defined on the boundary is related to the on-shell (Euclidean) action for
the solution in the bulk. Using our conventions one has
\be
{ F} = TS_E^{eff}\vert_{on~shell}
\ee
In the present case, after Wick rotating  action (\ref{efectiva}) and using the proposed ansatz one has
\begin{align}
{ S}_E^{eff}=\beta {L} \int dz &\left(-\frac{z}{2} \left(\phi'\right)^2+ \frac{z f(z)}{2} \left(A'\right)^2 - \frac{r_h^2 f(z)}{z} \left(\psi'\right)^2  - \frac{1}{z f(z)} \phi^2\psi^2 \right. \nonumber\\
&\left.+ \frac{1}{z} A^2 \psi^2 - \frac{r_h^2 m^2}{z^3} \psi^2 -  \kappa\left(\phi A' - A \phi'\right)
\right)
\label{last}
\end{align}
where Euclidean time has been integrated  and $L$ is the length of the $x$
interval.   Note that we have not included the purely gravitational last
term in (\ref{efectiva}) since being the metric a background,  it will
cancel out once  the difference of free energies between different phases
is considered.

After integrating by parts and using the equation of motion one gets from (\ref{last})
\begin{align}
T { S_\epsilon} = L \left[-\frac{z}{2} \phi' \phi + \frac{z\, f}{2} A' A - \frac{r_h^2 \,f}{z} \psi' \psi \right]_\epsilon^1 + L  \int_\epsilon^1 dz \left( \frac{\phi^2}{f} - A^2\right) \frac{\psi^2}{z}
\end{align}
where we have introduced a cutoff $\epsilon$ in order to handle the usual
action divergencies.  Using
the boundary conditions (\ref{Des})-(\ref{Dos}) one finds for the boundary term  {in the small $\kappa$ perturbative regime}
\begin{align}
T\, { S}_{ {\epsilon\,\rm boundary}} =   L  r_h^2 \left(D_1 D_2 + D^2_2\right) + \text{divergent terms in $\log\!\epsilon$}
\end{align}
Once   divergent terms are subtracted
 the $\epsilon \to 0$ limit can be safely taken and the
free energy per unit length   takes the form
\begin{align}
\frac{  F}{L} =-   \int_0^1 \!\! dz \left( \frac{\phi^2}{f} - A^2\right) \frac{\psi^2}{z} - r_h^2 \left(D_1 D_2 + D^2_2\right)
\label{cincuenta}
\end{align}

{ Using this formula we have calculated the free energy for the  $D_1=0$  solution  found in the previous Sec.ion. We find that  the free energy is negative definite in all the temperature range, opposite to what happens in the case $D_2=0$. These results are plotted in Fig. 3.}

~

\begin{center}
\includegraphics[width=4in]{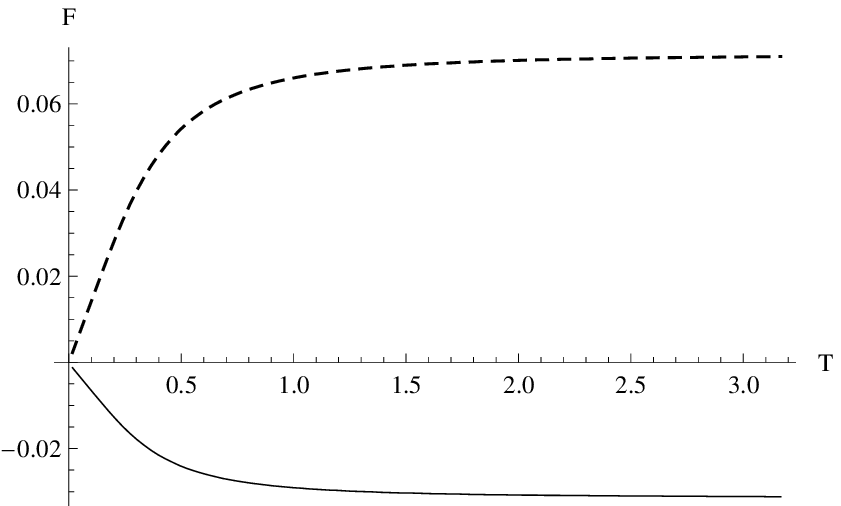}
\end{center}
~

 \noindent Figure 3: The free energy $F$  as a function of the temperature $T$ for the solution with $D_1=0$ (solid line) and $D_2=0$ (dashed line) for the same choice of parameters as in Fig. 1. The $D_2=0$ curve has positive free energy so is not thermodynamically favored.

\vspace{1 cm}


From Eq.(\ref{cincuenta})  one can see that the alternative $D_2 = 0$
choice of boundary conditions
causes the boundary contribution to the free energy to vanish. As for the
bulk contribution,  we have found numerically
that, in this case, it turns out to be positive definite in all the
temperature range (positive definite Euclidean action). This is due to the
$A^2$ negative term in eq.(\ref{cincuenta}) and directly related to the
necessity, when a Chern-Simons term is present, to include an $A_x$
component to solve eqs. (\ref{sistema}). In view of these results, only the
$D_1 = 0$ boundary condition should be considered when applying the AdS/CFT
correspondence to the theory defined by effective action (\ref{efectiva}).

\section{Summary and discussion}

The existence of a parity anomaly in odd-dimensional fermionic theories
coupled to gauge and gravitational fields and the consequent emergence of
Chern-Simons terms in the effective action induces a variety of phenomena
that are of interest both in high energy physics and in condensed matter
physics. Recent results of the application of the AdS/CFT correspondence to
study strongly coupled field theories using a weakly coupled gravity dual
motivated the investigation presented in this paper where an Abelian gauge
theory coupled to a complex scalar and massive fermions in an
asymptotically AdS$_3$ space-time was discussed. The main idea of our work
was to integrate out fermions in the path-integral defining the bulk
partition
function, so that finite temperature quantum fermion effects were incorporated in the   effective action. After solving the associated equations of motion, the AdS/CFT correspondence was applied to investigate the strong coupling behavior of the dual quantum theory defined in the boundary.

Although the explicit form of the effective action for pure gauge and pure
gravitational theories has been known for a long
time~\cite{Red}-\cite{Kulikov}, the case in
which both gauge and gravitational fields are present seems not to be
studied so the first step in our investigation was to determine whether
terms mixing gauge  and spin connections were induced by the parity anomaly.
Our  result, eq.(\ref{rhsfinal}) shows that no mixing term arises and the
odd-parity part of the effective action is just the addition of two
Chern-Simons terms, one for each connection. This result extends to the
finite temperature case leading to an affective action $S_E^{eff}$ which in
the weak coupling regime is given by eq.(\ref{efectiva}).

The next step in our analysis consisted in finding solutions to the equations of motion for $A_0, A_x$ and $\psi$ in an AdS$_3$ black hole background. Once the appropriate boundary conditions were imposed the solution can be used to saturate the bulk partition function and, using the gauge-gravity recipe, to analyze the fate of symmetry breaking in the dual quantum field theory. We worked in the probe limit taking as a gravitation background an AdS$_3$ black hole and the main result of our numerical analysis can be  summarized by Figs. 1 and 2. In the first one one can see that dual system does not show a phase transition to a normal (disordered) phase but remains in the ordered (symmetry breaking) state in the whole range of temperatures. One can clearly see that such result derives from the presence of the Chern-Simons term and will be present far all non-zero values of the gauge-fermion coupling constant. Concerning the free energy represented in figure 2, its behavior shows that the solution remains stable in the whole range of temperatures. Whether these results can be associated to the infinite  order phase transition found for example for Ising models defined on networks where the critical temperature becomes infinite \cite{Ale}-\cite{Do} would require  worthwhile analysis including the study of back reaction of matter fields on the geometry.

The fact that the presence of the fermions and their contribution to the effective action through the Chern-Simons term drastically changes the behavior of the system at the boundary could be seen as a striking result. One should however recall that the addition of a Chern-Simons term to a $2+1$ Maxwell-Einstein model radically changes the solution: without the Chern-Simons term the solution of the coupled system is a charged black-hole and a nontrivial gauge field potential $A_0$. Once the Chern-Simons term is introduced, no matter the strength of its coupling, the gauge field decouples and the black-hole becomes uncharged \cite{Banados}. We have numerically confirmed this result finding that the only solution in the uncondensed phase is the trivial one. There are other domains where the introduction of a Chern-Simons term completely change the properties of the theory. At the level of classical equations of motion it is well known that existence of Bogomol'nyi-Prasad-Sommerfield (BPS) equations for the Chern-Simons-Higgs model requires a sixth order symmetry breaking potential instead of the forth-order one required in the Maxwell-Higgs model. This implies the nonexistence of BPS equations for the complete Maxwell-Chern-Simons-Higgs system, no matter the strength of each coupling. The origin of these properties can be traced back to the relation between electric charged and magnetic field imposed by the presence of the Chern-Simons term. Through the connection between BPS equations and supersymmetry algebra, a similar condition hold concerning the possibility of supersymmetric extensions of these models \cite{Bogo}-\cite{Wein}.

\vspace{1cm}

\noindent\underline{Acknowledgments}: We thank Diego Correa, Carlos N\'u\~nez and Eduardo Fradkin for their helpful comments at different stages of our investigation. We also thank Jie Ren for his suggestions concerning the numerical computation.
We would like to thank the referee for his comments and suggestions that considerably improved the presentation of our results. This work was supported by  CONICET, ANPCYT,CIC, UNCuyo, UNLP, Argentina and NEU, USA.

 \end{document}